# FORMT: *F*orm-based *M*utation *T*esting of Logical Specifications

*Andreas Faatz, Andreas Zinnen*
*SAP AG*
*SAP Research CEC Darmstadt*
*Bleichstrasse 8*
*64287 Darmstadt*
*Germany*

## Abstract

*The draft paper defines a system, which is capable of maintaining bases of test cases for logical specifications. The specifications, which are subject to this system are transformed from their original shape in first-order logic to form-based expressions as originally introduced in logics of George Spencer-Brown. The innovation comes from the operations the system provides when injecting faults- so-called mutations - to the specifications. The system presented here applies to logical specifications from areas as different as programming, ontologies or hardware specifications.*

## Introduction/Motivation

Software testing knows the sub-discipline of so called fault-injection techniques. These techniques aim at maintaining a base of test cases for particular software. The assumption of these techniques is that programmers are competent and that a set of test cases is sensitive to any randomly produced change of the program code. Consider the following lines of pseudo-code:

```
Rem Snippet A;
Int i;
Input i;
If (i >= 0) print: "your input value is a non-negative number";
End.
```

The running program is intended to work in the following way. It would take an integer value and return the remark, that the input value is positive. Test cases for this program could for instance include

a) the passing of a fixed positive value, say 13 for the desired outcome:
`"your input value is a non-negative number"`

b) the passing of a fixed non-negative value, say -7 for the desired outcome: nothing or at least **not** "`your input value is a non-negative number`"

With the above lines of pseudo code compiled, running test case a) would return the remark that the input value is positive, whereas test case b) would return nothing, i.e. the program simply ends. Such a behavior would exactly match the expectations and intentions of the above program and the collection of test cases *seems* to cover the intended behavior. To illustrate fault-injection we assume a change of the operator >= to > ending up with

```
Rem Snippet B;
Input i;
If (i > 0) print: "your input value is a non-negative
number";
End.
```

Mathematically this is not the same and even not desirable, as passing 0 to the running program would return an inadequate void, i.e. would return nothing. The test cases must be extended to cope with this small change of the program code. This can be achieved by introducing a third test case

c) the passing of 0 for the desired outcome: "`your input value is a non-negative number`"

Then a successful total test run would return the desired output lines if a), b) and c) are run with Snippet A and the three test cases would be sensitively formulated, such that Snippet B can be distinguished from Snippet A, as c) would fail in the sense, that the desired outcome would not be reached.
This example illustrates the basic principles of mutation-based testing. Changes (so-called mutations) of the original code (origin) introduces mutants (in our case Snippet B would a mutant of Snippet A) and the base of test cases (in our example case a), b) and c)) must be able to detect a mutant. A general problem is the fact, that some mutants have exactly the same functionality as their origin. Consider for instance the following mutant, where the mutation is the injection of a NOT-operator and turning >= into =<:

```
Rem Snippet A';
Int i;
Input i;
If NOT(i < 0) print: "your input value is a non-negative
number";
End.
```

Snippet A' behaves exactly in the same way as Snippet A, i.e. no matter how the test base would look like, each test case would return the same outcome from Snippet A and snippet A' respectively. Mutations like the ones from Snippet A to B are called *true mutations*. B is called a *true mutant* of A, in contrast to A', which is not a true mutant of A. The ideal maintenance of a base of test cases would be – from the mutation-based testing perspective – that each true mutation can be detected by at least one of the test cases. The system called "FORMT" to be described in this paper describes a clean way of generating true mutations/mutants for logical specifications. The draft is organized as follows: we start with a problem statement, introduce related work and the relevant background of Spencer Brown's meta-theory. Then we restrict the problem space ("requirements and scope") and give a description of FORMT. We also summarize the main achievements and give a perspective on future work.

## Problem addressed by FORMT

Many formal ways of expressing a model or specification are using formal logic. Prominent examples for the use of first-order logic are (i) the specification of workflows, which can be based on a sound language-independent foundation using Petri Nets [Aa98]. Petri Nets themselves can be expressed by expressions in first order logic [So00], (ii) declarative programming by means such as the languages Prolog [Br88] or (iii) specification languages such as Z [ISO02]. Besides these more software-oriented applications, (iv) the specification of hardware by expressions in first-order logic is common [REH99]. Another example for the notation of rules by first-order logic in computer science are (v) the concepts contained in and the rules applying to ontologies [GoFe04].
All these very different application areas create (potentially huge) code, which must be maintained and tested. Similarly to our algebraic example in the introduction/motivation, the generation (*ex ante*) or detection (*ex post*) of true mutants for logical specifications is not straightforward.
The system presented in this paper is called FORMT. It is a novel way to generate true mutations for particular types of logical specifications with applications in programming, ontologies and hardware specification, i.e. areas with heavy use of propositional logic or Boolean algebra and monadic predicate calculus. Moreover, it is also a novel way to browse the set of mutants and to let a user of FORMT determine, which parts of the mutated specification are not clearly covered by the tests contained in the test base.

## Closest related work

Generally speaking, the detection of true mutants *after* these have been generated (ex post approach) may yield to non-decidable situations [OfPa97]. Thus, any restriction of this problem brings significant progress.  FORMT presents a way to generate true mutants for a *particular* novel mutation

operation. Thus, FORMT is a shift away from checking true mutation ex post strategies like applied

- by [HDD99], who slice program code, for example nested `if`-clauses to make the detection of mutations computationally feasible
- weak mutation testing, which evaluates components of programs along with their respective (from the total application perspective: internal) states [Ho82]

Instead, FORMT is an ex ante strategy for a particular set of novel mutation operators. This set is based on the form-based logic calculus [Sp72], also called calculus of distinctions. To our knowledge, this calculus was not applied to any mutation generation or checking until now at all. Re-translating the operations from FORMT to classical first-order logic also shows, that the mutation operators in FORMT are **not** a re-formulation of the ones introduced by [BOY00]. The operation of FORMT, i.e. switching of the form expression, is novel.

FORMT also differs from other ex ante strategies, namely from

- [AHH04], who apply Genetic Algorithms to evolve true mutants. This is a randomized approach, whereas the core mechanisms of FORMT are deterministic.
- selective mutation [ORZ93], which just allows subsets of the mutants to be produced. FORMT differs from selective mutation, which makes selections by restricting the set of mutation operators. In contrast to that, FORMT consequently uses one mutation operator and checks, at which places of the specification the application of this operator takes place.
- schema-based mutation [UOH93] – although schema-based mutation compiles all potential mutants into one meta-level expression, unlike FORMT it is not dealing with logical specifications and not at all applying Spencer-Brown's form-based logical calculus

The aforementioned approaches have been applied to procedural and object-oriented programs. Although their principles would also hold for logical specifications, the state of the art mutation testing systems for logical specifications [BOY00], [OBY04] is based on mutation operators for model checking approaches. Still the detection of true mutants remains a problem. [SCS+03] is an example from the direction of mutation testing systems for logical IT-security specifications, which even goes as far as manually checking for true mutation. The approaches of mutation testing systems for logical specifications are able to

(i) automatically generate counter-examples, i.e. to automatically extend the test base in such a way, that a mutant is detected and

(ii)  give information about the current coverage of the test base, i.e. metrics on how strongly the test based must be evolved (by manual or automated test generation) to be able to discover mutants.

(i) is out of the focus of this paper, i.e. we abstract from the question, if the update of the test base happens manually or automatically. (ii) is presented as tables for different mutation operators in all related approaches. FORMT takes another approach by presenting results on test base coverage graphically and navigable. In contrast to [OBY04], the mutation happens by exactly one (and novel) type of operation.

## Existing background techniques of FORMT

We present the basic set-up and results of [Sp72] as we need them for the background techniques of the FORMT as a system. This also relates to the requirements on the logical specifications operable by FORMT. The *form-based calculus* is based on one operator which we denote with bold brackets as

**()**.

It is called "the form". This operator can be arranged in two ways:

**()()**, which means writing it sequentially, i.e. concatenating it,

and

**(())**, which means nesting it.

Based on this arrangements along with their asymmetry, the form-based calculus has two axioms. Expressed in the notation the axioms are:

**(())** =

and

**()()** = **()**.

That means nesting of the form makes it disappear and concatenation of the form reduces to one form. An arbitrary arrangement of forms is expressed as *a*, i.e. in small italics. If *a* is concatenated with a form, we write

*a***()**

or

**()**a.

Generally speaking, if two expressions *a* and *b* are concatenated we write

*ab*.

If *a* is nested into a form we write

**(**a**)**.

The form-based calculus allows for nesting and appending the basic expressions achieving for example

**()()()**,
**((()))**,
**((**a**))** or
**((**a**)()**b**)**.

Along with the two axioms and the commutativity *ab* = *ba*, we may simplify expressions, for instance:

**()()()** = **()**,

**((()))** = **()**,

**((**a**))** = *a*,

**((**a**)()**b**)** = **((**a**)())** = **(())** = .

[Sp72] also showed, how *some* of the expressions of first-order logic can be transformed into the form-based calculus. If we write **AND** for the logical conjunction, **OR** for the logical disjunction, **NOT** for logical negation and **=>** for logical implication, then we find the following translations:

 *a* **AND** *b* translates to **((**a**)(**b**))**

*a* **OR** *b* translates to *ab*

**NOT** *b* translates to **(**b**)**

*a* **=>**b translates to **(**a**)**b .

[Sp72] states, that the form-based calculus is capturing for such universal expressions well and that there is an extra effort to translate logical **EXISTS** quantors. Nevertheless, a sound translation of the **EXISTS** quantor cannot be found without extending the notations fundamentally. The logical **FORALL** quantor is translated to an implication. That means a translation to expressions like

**FORALL** *x*: *a(x)* **=>** *b(x)* translate to **(***a***)***b*.

In other words: the form-based calculus is equivalent to monadic predicate logic, where all predicates have only one argument. This fact will shape our requirements on applying FORMT.

## Requirements and scope of FORMT

The translation rules from the last section provide the scope of the paper. FORMT is designed for applications of Boolean algebra, for example electronic circuit design, branching conditions (if-then-else) in methods of an object-program or other computer program, relational database queries or (parts of) specifications in description logics, which use only or can be encapsulated/abstracted to structures using only **AND**, **OR**, **NOT** and implications (**=>**). Moreover a collection of tests (setting parameters of the specification, i.e. the *a* and *b* from the previous section) with the expected results (true or false in the case of setting all parameters or formulae in case of setting some parameters) is given. We claim FORMT to be able to process this in a novel way. Moreover, any specification resulting from monadic predicate logic with a respective test base as above can be processed by FORMT.
Finally, consider the case that the (still pending) proof of a translation from the full first order logic including **EXISTS** and **ALL** quantors as well as all of their combinations to the form calculus or a necessary extension will be achieved. We claim for this case, that the mutation mechanisms and the visualization principles of FORMT will also hold either

- for the translation to a form-based expression itself or
- for the parts of the expression, which are form-based in the sense of the previous notation subsection, i.e. in the sense of [Sp72].

In the latter more general case of the completed proof, FORMT would be able to process all specifications based on Petri-Nets, declarative programming and all rule-based systems using first-order logic, for instance the rules specified for an ontology.

## Description of the system FORMT

We explain and illustrate the core idea now, first in an abstract way, then by example. The example given in this section extends one of the examples given in appendix 2 of [Sp72]. We assume along with our previous section on requirements, that an adequate logical specification (i.e. a specification adequate for automated processing) and a test base are given.

*Step 1*

FORMT starts with the translation of the logical specification to a form-based expression. The translation was shown in the section on existing background techniques.

*Step 1b*
This form-based specification might be systematically simplified by the algebraic operations in [Sp72] – or just by the two laws **(())=**  and  **()()=()**.

*Step 2*

The resulting form-based expression will be sent through one of the following mutation steps creating two variants of FORMT:

- either by deletion of exactly one of the forms from the specification, i.e. in our notation from the previous section deleting a bracket **(** and its counterpart **)** – (this is *variant 2a* or *step 2a*),
- or  by adding a form to the specification, i.e. a bracket **(** and its counterpart **)** around one existing expression, be it composed by parameters *a*, *b* et cetera or existing concatenation of forms (this is  *variant 2b* or *step 2b*)

The latter adding step expressed by the second main bullet would be step 2b of FORMT. Steps 2a and 2b both result in a mutated specification. It is a true mutant. Repeating steps 2a and 2b (one has to decide for either variant, adding or deleting) with the logical specification from the start (not with the mutated one) generates a set of true mutants. For variant 2a there is a terminating condition, which is fulfilled if each form was deleted from the original translated form-based specification once. Both 2a and 2b are novel ways of creating true mutants.

*Step 3*

Step 3 tests all the mutants by instantiating one or all of the parameters in the mutated specifications from 2a or 2b; instantiating a parameter to **()** is the translation of setting it to **TRUE**, setting a parameter to empty - i.e. the equivalent of **(())** – means setting it to **FALSE**. A test base kills a mutant, if at least one of the results foreseen for the test, i.e. with the translation of **TRUE**, the translations of **FALSE** fails for the test parameters. Each mutant **M** can be assigned with information **I M** (e.g. a metrics) resulting from the set of tests, which fail or do not fail. An example metrics can be the percentage of tests failing.

*Step 4*

Step 4 of FORMT is visualization. Basically, the original logical specification translated in step 1 is visualized in the following way:

- the form **()** is presented as a circle or any other closed shape without self-crossing boundary (for example ellipses, rectangles et cetera), for instance

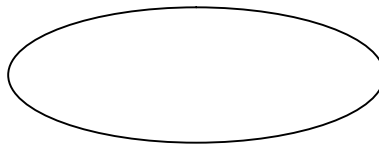

- nesting i.e. **(())** is presented as writing any closed shape, for instance a circle, into another shape representing the form, for instance

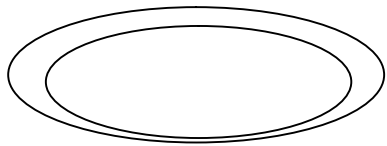

- concatenation i.e. **()()** is visualized by placing shapes next to each other without overlap for example

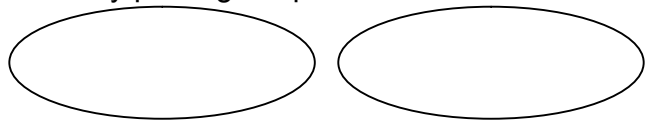

- parameters are written inside the shapes – see example for a visualization of

The innovation of the consequent step 4 is the presentation of information **I M** for each mutant **M**. In case of the branch resulting from 2a, **I M** will be written directly to, on or at the shape, which is deleted for true mutation. If for instance the test base fails to kill a mutant, this could be indicated by coloring the shape, where a deletion of the respective form takes place. In case of the branch resulting from 2b, **I M** can for example be inserted by a colored dotted line. And the dotted line is placed as a shape around the shape for the expression, where the mutant is produced by adding a form. Step 4 produces a novel comprehensive view on a set of true mutations, for which the test based should be extended, be it automatically or manually. Optionally, step 4 can be enhanced by partial or complete re-translations of the form-based expressions into classic propositional or predicate expressions.

We now give an example of the described steps of the system FORMT. Let the following logical specification (expression) be given:

*((p =>q)* **AND** *(r => s)* **AND** *(q* **OR** *s))* => *(q* **OR** *v)*

- Step 1 (translation): the specification translates to **((((p) q) ((r) s) (q s)))** *pr*, which can be simplified to

- Step 1b (simplification): **(qs)pr** [Sp72] or more directly to **((p) q) ((r) s) (q s)** *pr*. We'll refer to the simple **(qs)pr** for the continuation of the example.

- Step 2a (mutation by deletion): **(qs)pr** produces a true mutant, as there is exactly one form, that can be left out: *qspr* as well as the alternative

- Step 2b (mutation by adding a form): **((qs))pr** = *qspr* would be the equivalent mutation by adding a form.

- Step 3 (applying the test base, assigning information to the form-based expression): assume, the test base does not kill *qspr* and we attach this information to the first form of **(qs)pr**, i.e. the **(qs)**-part of the expression. The procedure for 2b is analogous, the information is "remembered" by the outer brackets of the **((qs))**-part in **((qs))pr**.

- Step 4 (visualization): The visualization result for 2a would be

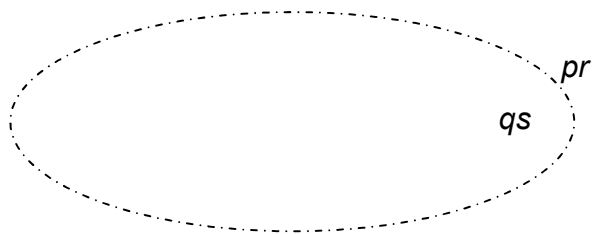

with the dashed line representing additional information, in this case a mutant not killed.

For 2b the representation is

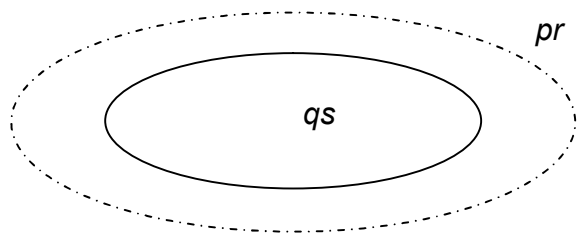

with the dashed line representing a mutant not killed and the continuous line representing the initial form in the expression **(*qs*)***pr*.

The diagrams could be partially re-translated into conventional logic expressions, for example *qs* to *(q* **OR** *s)*. The result is a map of true facilitating the analysis of the test base, for example detecting, which variables and partial expressions cause the test base to fail in the sense of not killing the mutant. Consider especially the case, where instead of two or one form many (for example thousands) forms occur with embeddings and concatenations. FORMT as a system would be implemented with a browsing facility allowing for instance for zooming. Moreover, there are different possibilities for enhancing the layout by grouping criteria, which also can be mixed:

- grouping forms with similar variables together
- grouping forms with a similar depth, i.e. steps of nesting, together
- grouping killed mutants and mutants not killed to different sectors
- grouping forms according to how many killed or not killed mutants are resulting from a nested form

Finally, also dashed lines were just an example. The information available for a mutant could also influence the shape itself, for instance making it a rectangle for mutants not killed and leaving it oval for the other ones. Also coloring would be an option. Combinations of these options would refer to different information visualized for the "killing" behavior of the test base.

## Conclusion

At the stage of this draft paper, FORMT claims to be novel with respect to the following core achievements:

- FORMT is a novel method of generating mutants for logical specifications. It is the first system applying form-based expressions in mutation-based testing.
- FORMT is a novel visualization and browsing technique for mutated logical specifications against a test base. It is the first system using form-based expressions as a basis for a comprehensive view on the mutated logical specification and on its behavior with respect to being killed by a test base or other information a mutant might bare with respect to the test base.

## Outlook

The draft paper presented an approach to testing specifications, which are using shapes from the classic Boolean approach. At this stage we point at a type of specifications, which is not covered by our investigations yet but would be a candidate for future work: specifications *originally written* by means of the form, including its capabilities of self-reference. An example of frameworks for such originally form-oriented specifications can be found in the sociologically motivated "studies of the next society" by Baecker [Bae07], who proposes the form as means of specification. In such specifications, the form and its axiomatically possible patterns provide a description how e.g. educational and scientific institutions, families, teams and companies work culturally, i.e. draw their distinctions on the background of disruptive information shifts as introduced by new media available to a society. Among the interesting resulting questions we identify: how would a test base and a mutation look like – and what are the impacts of introducing tests and mutations for the *scientific* narration behind the specification?

## APPENDIX A: Justification of the *ex ante* mutation operation of FORMT

Spencer Brown [Sp72] distinguishes so-called transparent and opaque spaces, i.e. form-based expressions, which oscillate between **TRUE** and **FALSE** or not with the change of variables **TRUE** or **FALSE**. A form-based expressions only becomes independent from change of **TRUE** or **FALSE**, if it contains empty forms, for instance if it looks like **()***a*. This is not possible by the translation rules for the logic expressions, i.e. all spaces here are transparent, i.e. any deletion or adding of a form will have valuable consequences for the form-based expression around the position, where a mutant is generated by adding or deleting a form. Thus we conclude that the operations described for our FORMT create true mutants.